
\input phyzzx
\vsize 9.25in
\hsize 6.3in

\rightline{UATP-95/02}
\rightline{March 1995}
\vskip 0.2in
\centerline{\seventeenbf Soliton Induced Singularities}
\centerline{\seventeenbf in 2 d Gravity and their Evaporation}
\vskip 0.75in
\centerline{\caps Cenalo Vaz\footnote{\dagger}{Internet:
cvaz@mozart.si.ualg.pt}}
\centerline{\it Unidade de Ci\^encias Exactas e Humanas}
\centerline{\it Universidade do Algarve}
\centerline{\it Campus de Gambelas, P-8000 Faro, Portugal}
\vskip 0.2in
\centerline{and}
\vskip 0.2in
\centerline{\caps Louis Witten\footnote{\dagger\dagger}{Internet:
witten@ucbeh.san.uc.edu}}
\centerline{\it Department of Physics}
\centerline{\it University of Cincinnati}
\centerline{\it Cincinnati, OH 45221-0011, U.S.A.}
\vskip 0.75in
\centerline{\bf \caps Abstract}
\vskip 0.2in

Positive energy singularities induced by Sine-Gordon solitons in
1+1 dimensional dilaton gravity with positive and negative
cosmological constant are considered. When the cosmological
constant is positive, the singularities combine a white hole, a
timelike singularity and a black hole joined smoothly near the
soliton center. When the cosmological constant is negative, the
solutions describe two timelike singularities joined smoothly near
the soliton center. We describe these spacetimes and examine their
evaporation in the one loop approximation.
\vfill
\eject

Since the pioneering work of Hawking${}^1$ and others${}^2$ in the
mid-seventies, the classical and quantum analysis of singularities
in gravity has led to a lively discussion${}^{3,4}$ on some
fundamental physical problems that are expected to arise in an
eventual theory of quantum gravity. Hawking's discovery that black
holes evaporate thermally raised a deep puzzle in physics, the
possibility of information loss in quantum gravity. Attempts to
solve this problem have characteristically been hampered by
technical difficulties, particularly the perturbative non-
renormalizability of quantum gravity.

Not long ago, Witten and others${}^5$ discovered a 1+1 dimensional
model of gravity with non-trivial dynamics which contains many of
the key features of the four dimensional theory, in particular the
formation and evaporation of black holes and naked singularities.
As the most interesting features of the four dimensional theory are
retained in this model and the dynamics is considerably simpler
than in four dimensions, one has a potentially useful device to use
in examining some of the interesting physics underlying the
problems posed by Hawking's early work. With this motivation, a lot
of effort has been directed toward understanding 1+1 dimensional
dilaton gravity.

An early attempt at understanding the dynamical formation and
evaporation of black holes in 1+1 dimensions appeared in the work
of Callan, Giddings, Harvey and Strominger (CGHS)${}^6$ in which
the authors coupled matter degrees of freedom to the original
Witten model by way of conformally invariant scalar fields whose
solution was taken to be a shock-wave travelling at constant
advanced time.  Neglecting the back reaction, the incoming shock-
wave was seen to radiate away  before the black hole forms but,
because the dilaton coupling is not small at the turn around point,
the back reaction becomes important and the one loop approximation
is an unreliable indicator of what is actually happening. Various
improvements have since been made to the original CGHS
model${}^{7,8}$, but in some form or other they have all had their
failures and a reliable understanding even of two dimensional black
hole dynamics is yet to come (for a review on the status of the
CGHS and other related models, see [9]). Naked singularities were
also analyzed within the context of the CGHS model with a negative
cosmological constant.${}^{10}$ In the one loop approximation,
neglecting the back reaction, the naked singularity evaporates
catastrophically emitting all its energy at early retarded times.
It was argued that for energetic shock waves the dilaton coupling
is weak in the evaporation region (at the ``explosion'' point)
making the one loop approximation a good indicator of the
underlying physics.

Recently a model of 1+1 dimensional dilaton gravity with Sine-
Gordon solitons${}^{11}$ was proposed as being of interest from the
point of view of non-linear integrable systems. The model may also
be interesting from the point of view of soliton solutions in four
dimensional general relativity for spacetimes admitting an Abelian
two parameter group of isometries.${}^{12}$ It lacks conformal
invariance in the matter sector, but one hopes that the model's
integrability will eventually facilitate its full canonical
quantization.  The singularities induced by the incoming solitons
are qualitatively quite different from those induced by the shock
wave of the CGHS model. They are made up of spacelike and timelike
pieces joined smoothly by lightlike singularities at the soliton
center. Thus the model combines black holes and naked singularities
and one might ask what quantum effects do to these objects and what
is the ultimate fate of the incoming soliton, taking into account
its Hawking radiation.  (As the classical stress energy falls off
exponentially on ${\cal I}^+$, the dominant effect there is its
Hawking evaporation.) As a step towards answering this question, we
briefly describe the classical solutions, showing that they do
indeed describe the dynamical formation of positive energy
singularities, and then examine their evaporation in the one loop
approximation. It turns out remarkably that the Hawking radiation
does not differ significantly from its shock wave counterpart if
natural boundary conditions are imposed.

We consider the CGHS action modified to include the Sine-Gordon
potential
$$S~~ =~~ {1 \over {2\pi}}~ \int d^2 x \sqrt{-g} \left[ e^{-2\phi}
\left( -R~ +~ 4 (\nabla \phi)^2~ +~ \Lambda\right)~ -~ {1 \over 2}
(\nabla f)^2~ +~ 4 \mu^2 e^{-2\phi} (\cos f~ -~ 1) \right]
\eqno(1)$$
where $\phi$ is the dilaton, $f$ is a matter field and $\Lambda$ is
the cosmological constant which we will take to be either positive
or negative in what follows. $R$ is the two dimensional scalar
curvature, and $g_{\mu\nu}$ is the two metric. Our conventions are
those of Weinberg.${}^{13}$ This action without the $f-$matter
fields arises in two dimensional string theory.

The classical equations of motion follow by variation with respect
to the metric, dilaton and matter fields. The metric equations
$$\eqalign{0~ =~ {\cal T}_{\mu \nu}~ &=~ e^{-2\phi}\left[ 2
\nabla_\mu \nabla_\nu \phi~ -~ {1 \over 2} e^{2\phi} \nabla_\mu f
\nabla_\nu f \right. \cr&+~ \left. g_{\mu\nu} \left( 2 (\nabla
\phi)^2~ -~ 2 \nabla^2 \phi~ -~ \Lambda~ +~ {1 \over 4} e^{2\phi}
(\nabla f)^2~ -~ 2 \mu^2 (\cos f~ -~ 1) \right)
\right]\cr}\eqno(2)$$
form a set of constraints on the allowable field configurations.
The dilaton and matter equations are
$$4 \nabla^2 \phi~ -~ 4 (\nabla \phi)^2~ -~ R~ +~ 4 \left( {\Lambda
\over 4}~ +~ \mu^2 (\cos f~ -~ 1) \right)~~ =~~ 0\eqno(3a)$$
and
$$-\nabla^2 f~ +~ 4 \mu^2 e^{-2 \phi} \sin f~~ =~~ 0\eqno(3b)$$
It is simplest to analyze the above equations in the conformal
gauge, where the metric has the form
$$g_{\mu\nu}~~ =~~ e^{2 \rho} \eta_{\mu\nu}, \eqno(4)$$
and in lightcone coordinates $x^\pm~ =~ x^0 \pm x^1$ which we use
hereafter. The constraints and equations of motion reduce to
$$\eqalign{0~~ &=~~ {\cal T}_{++}~~ =~~ e^{-2\phi} \left[ -4
\partial_+ \rho \partial_+ \phi~ +~ 2 \partial_+^2 \phi \right]~ -~
{1 \over 2} (\partial_+ f)^2\cr0~~ &=~~ {\cal T}_{--}~~ =~~ e^{-
2\phi} \left[-4 \partial_- \rho \partial_- \phi~ +~ 2 \partial_-^2
\phi \right]~ -~ {1 \over 2} (\partial_- f)^2\cr 0~~ &=~~ {\cal
T}_{+-}~~ =~~ e^{-2\phi} \left[-2 \partial_+ \partial_- \phi~ +~ 4
\partial _+ \phi \partial_- \phi~ +~ {\Lambda \over 4} e^{2 \rho}~
+~ \mu^2 e^{2 \rho} (\cos f~ -~ 1) \right]\cr} \eqno(5)$$
and
$$\eqalign{&-4 \partial_+ \partial_- \phi~ +~ 4 \partial_+ \phi
\partial_- \phi~ +~ 2 \partial_+ \partial_- \rho~ +~e^{2 \rho}
\left[{\Lambda \over 4}~ +~ \mu^2 (\cos f~ -~ 1) \right]~~ =~~ 0\cr
& +\partial_+ \partial_- f~~ +~~\mu^2 e^{2(\rho-\phi)} \sin f~~ =~~
0\cr} \eqno(6)$$
When the dilaton equation in (6) is combined with the last of the
three constraints in (5), one sees that the conformal factor,
$\rho(x)$, is equal to the dilaton, $\phi(x)$ up to a harmonic
function, $h(x)$. However, a choice of $h(x)$ is essentially a
choice of coordinate system because the choice of conformal gauge
does not fix the conformal subgroup of diffeomorphisms. Choosing a
coordinate system such that $h(x)=0$, the general solution,
satisfying the constraints has the form ($e^{-2 \rho} = e^{-2 \phi}
= \sigma$)
$$\eqalign{f_{kink}~~ &=~~ 4 \tan^{-1} e^{(\Delta~ -~ \Delta_0)}\cr
\sigma~~ &=~~ a~ +~ bx^+~ +~ cx^-~ -~ {\Lambda \over 4} x^+ x^-~ -~
2 \ln \cosh(\Delta~ -~ \Delta_0)\cr} \eqno(7)$$
in terms of $\Delta = \gamma_+ x^+~ +~ \gamma_- x^-$, where
$$\gamma_\pm~~ =~~ \pm~ \mu \sqrt{{1 \pm v} \over {1 \mp v}},
\eqno(8)$$
$v$ is the velocity of the soliton, $f(x,t) = f(x+vt)$, $\Delta =
\Delta_0$ is its center which we take without loss of generality to
be greater than or equal to zero, and $a$, $b$ and $c$ are
arbitrary constants. The classical energy momentum tensor of the
incoming soliton is
$$\eqalign{T^f_{++}~~ &=~~ {1 \over 2} (\partial_+ f)^2~~ =~~ {{4
\gamma_+^2} \over {\cosh^2(\Delta - \Delta_0)}}\cr T^f_{--}~~ &=~~
{1 \over 2} (\partial_- f)^2~~ =~~ {{4 \gamma_-^2} \over
{\cosh^2(\Delta - \Delta_0)}}\cr T^f_{+-}~~ &=~~ \mu^2 (\cos f~ -~
1)~~ =~~ -~ {{2 \mu^2} \over {\cosh^2(\Delta - \Delta_0)}}. \cr}
\eqno(9)$$
To fix the constants of integration one needs to impose some
reasonable physical conditions. Thus, we require that the metric
reduces to the linear dilaton vacuum in the absence of the incoming
soliton, which we define as the limit $T^f_{\mu\nu} \rightarrow 0$,
$\Delta_0 = 0$. This gives
$$\sigma~~ =~~ -~ {\Lambda \over 4} x^+ x^-~ -~ 2 \ln \cosh(\Delta~
-~ \Delta_0) \eqno(10)$$
There is also the antikink solution
$$\eqalign{f_{antikink}~~ &=~~ 4 \arctan e^{-(\Delta~ -~ \Delta_0)}
\cr \sigma~~ &=~~ -~ {\Lambda \over 4} x^+ x^-~ -~ 2 \ln \cosh
(\Delta~ -~ \Delta_0)\cr} \eqno(11)$$
which may be analyzed analogously. We will work only with the kink
solution in (7) and (10).  When $\Lambda > 0$ equation (10)
represents a spacetime that admits a positive energy singularity
combining a white hole, a timelike singularity and a black hole,
all smoothly joined along the soliton center (by a white hole we
mean a spacelike naked singularity). When the cosmological constant
is negative, the spacetime admits two naked singularities joined
smoothly at the soliton center.  In every case, the curvature
singularity is at $\sigma = 0$ as can be seen by inserting (10)
into the expression
$$\eqalign{R~~ &=~~ +~ 2 e^{-3\rho} \nabla^2 e^\rho~~ -~~ 2 e^{-
4\rho} (\nabla e^\rho)^2 \cr &=~~ +~ 4~ \sigma~ \partial_+
\partial_- \ln \sigma \cr} \eqno(12)$$
for the curvature scalar. The Bondi energy${}^{14}$ of the
singularity on ${\cal I}^+$ is straightforward to calculate in each
case, as the spacetime admits a Killing vector near null infinity.
For example, near ${\cal I}_R^+$ and for positive cosmological
constant, the metric behaves as
$$\sigma~~ \rightarrow~~ -~ \lambda^2 x^{+'} x^{-'}~~ +~~
2\Delta_0~ -~ {{4 \mu^2} \over {\lambda^2}}~ +~ 2 \ln 2 \eqno(13)$$
where
$$\eqalign{x^{+'}~~ &=~~ x^+~ +~ {{2 \gamma_-} \over
{\lambda^2}}\cr x^{-'}~~ &=~~ x^-~ +~ {{2 \gamma_+} \over
{\lambda^2}}\cr}\eqno(14)$$
The (timelike) Killing vector is $\xi^\mu = (x^{+'}, x^{-'})$. Let
$t_{\mu\nu}$ be a linearization of ${\cal T}_{\mu\nu}$ about the
(dilaton) vacuum so that $j_\mu = t_{\mu\nu} \xi^\nu$ is a
conserved current in the asymptotic region. Consider a solution
that is asymptotic to the vacuum with $\phi = \phi^{(0)} + \delta
\phi$, where $\phi^{(0)} = - \ln(- \lambda^2 x^{+'} x^{-'}) /2$.
The current density takes the form
$$\eqalign{j_{+}~~ &=~~ 2 \lambda \partial_{+'} \left( e^{-
2\phi^{(0)}} \left[ \delta \phi~ +~ x^{+'} \partial_{+'} \delta
\phi~ +~ x^{-'} \partial_{-'} \delta \phi \right] \right) \cr j_{-
}~~ &=~~ 2 \lambda \partial_{-'} \left( e^{-2\phi^{(0)}} \left[
\delta \phi~ +~ x^{+'} \partial_{+'} \delta \phi~ +~ x^{-'}
\partial_{-'} \delta \phi \right] \right). \cr} \eqno(15)$$
The conservation of $j_\mu$ implies the existence of two charges,
$Q^+ = \int^{{\cal I}_R^-} dx^- j_-$ and $Q^- = \int^{{\cal I}_R^+}
dx^+ j_+$ which evolve the system in the direction of increasing
$x^+$ and $x^-$ respectively and close to infinity. The current
density is a total derivative, so its integral can be measured as
a surface term. Thus, for example, one obtains the conserved charge
(the Bondi energy)
$$\eqalign{M~~ =~~ Q^-~~ &=~~ 2 \lambda \left( e^{-2\phi^{(0)}}
\left[ \delta \phi~ +~ x^{+'} \partial_{+'} \delta \phi~ +~ x^{-'}
\partial_{-'} \delta \phi \right] \right)_{{\cal I}_R^+}\cr}
\eqno(16)$$
on ${\cal I}_R^+$.

If $\Lambda = 4\lambda^2 > 0$ the spacetime described by
$$\sigma~~ =~~ -~ \lambda^2 x^+ x^-~ -~ 2 \ln \cosh(\Delta~ -~
\Delta_0) \eqno(17)$$
is a combination of a white hole, a black hole, and a naked
singularity. The Kruskal diagram displayed in figure I shows the
singularity along with the trajectory of the soliton center.
Directing our attention to the region on right, the observer on
${\cal I}_R^+$ measures the (Bondi) mass
$$M_R~~ =~~ 4\lambda \left(\ln 2~ -~ {{2\mu^2} \over {\lambda^2}}~
+~ \Delta_0 \right).\eqno(18)$$
and the soliton center is seen to emerge from the merging of a
white hole and a timelike singularity at $(x^+ = 0, x^- = \Delta_0
/ \gamma_-)$. The white hole extends from $(x^+=2 \gamma_-
/\lambda^2, x^-=-\infty)$ on ${\cal I}_R^-$ to $(x^+ = 0, x^- =
\Delta_0/\gamma_-)$.  Here it smoothly turns into a timelike line
proceeding to $(x^- = 0, x^+ = \Delta_0/ \gamma_+)$, where the
soliton is reabsorbed. At this point the singularity once again
turns spacelike smoothly and reaches ${\cal I}_R^+$ at $(x^+ =
\infty, x^- = - 2 \gamma_+/\lambda^2)$.
All singularities are asymptotically spacelike as they approach
${\cal I}^\pm$. Although the figure was drawn for a particular
choice of parameters, the qualitative behavior of the singularities
is independent of the choice as long as the Bondi mass is positive.

The soliton never enters the left region (we have taken $\Delta_0
\geq 0$) and the observer on ${\cal I}_L^+$ measures the mass
$$M_L~~ =~~ 4\lambda \left(\ln 2~ -~ {{2\mu^2} \over {\lambda^2}}~
-~ \Delta_0 \right)\eqno(19)$$
If $M_R$ and $M_L$ are both positive, the singularities on the left
and on the right have the same features.  As $\Delta_0 \rightarrow
0$ (figure II) the left and right singularities merge, in the limit
forming a white hole that extends from $x^+ = -2\gamma_-/\lambda^2$
on ${\cal I}_R^-$ to $x^- = 2\gamma_+ / \lambda^2$ on ${\cal I}_L^-
$ and a black hole that stretches from $x^- = -2\gamma_+
/\lambda^2$ on ${\cal I}_R^+$ to $x^+ = 2\gamma_- / \lambda^2$ on
${\cal I}_L^+$, intersecting at the origin. The length of the
timelike naked singularity has shrunk to zero and the masses
measured on both null infinities are now the same. Even if they are
zero ($2 \mu^2 = \lambda^2 \ln 2 $) soliton energy and momentum is
present throughout the spacetime.

The spacetime with negative cosmological constant ($\Lambda = - 4
\lambda^2$)
$$\sigma~~ =~~ \lambda^2 x^+ x^-~ -~ 2 \ln \cosh(\Delta~ -~
\Delta_0) \eqno(20)$$
is shown in figure III. All singularities are timelike as they
approach ${\cal I}^\pm$. In the top region the timelike singularity
approaches ${\cal I}_R^+$ at ($x^+ = \infty, x^- = 2\gamma_+/
\lambda^2)$ and approaches ${\cal I}_L^+$ at $(x^+ = - 2 \gamma_-
/ \lambda^2, x^- = \infty)$. The two timelike sections merge at a
white hole in the region where the soliton center enters the
spacetime. The latter emerges at $(x^+ = \Delta_0 / \gamma_+, x^-
= 0)$ and travels to $i^0$. The Bondi mass depends on whether the
asymptotic observer is located on ${\cal I}_R^+$ or ${\cal I}_L^+$,
$$\eqalign{M_R~~ &=~~ 4\lambda \left({{2\mu^2} \over {\lambda^2}}~
+~ \Delta_0~ +~ \ln 2 \right)\cr {\rm and}~~~~~ M_L~~ &=~~ 4\lambda
\left({{2\mu^2} \over {\lambda^2}}~ -~ \Delta_0~ +~ \ln 2
\right),\cr} \eqno(21)$$
the qualitative behavior again being independent of the parameter
values chosen if both masses are positive. The masses are equal
when $\Delta_0 = 0$. In this limit one has two naked singularities,
the first extending from $x^- = - 2 \gamma_+/ \lambda^2$ on ${\cal
I}_L^-$ to $x^- = 2 \gamma_+/ \lambda^2$ on ${\cal I}_R^+$ and the
other from $x^+ = 2 \gamma_-/ \lambda^2$ on ${\cal I}_R^-$ to $x^+
= - 2 \gamma_-/ \lambda^2$ on ${\cal I}_L^+$, intersecting at the
origin (figure IV).

We would now like to include quantum effects in this model.  As
seen from (9), the classical stress energy tensor is exponentially
vanishing on ${\cal I}_R^+$ ($x^+ = \infty$) so that the Hawking
evaporation is dominant here. To include quantum effects in two
dimensions one needs only the trace of the stress tensor, the other
components being determined by the conservation equations in
keeping with Wald's axioms.${}^{15}$ The geometric contribution to
the trace can depend only on the scalar curvature, $R$, as it is
the only available geometric invariant. Thus the quantum correction
to the trace of the stress tensor must be given by
$${T^{(q)\mu}}_\mu~~ =~~ -~ 4 \sigma T^q_{+-}~~ =~~ -~ \alpha R~~
=~~ -~ 4 \alpha \sigma \partial_+ \partial_- \ln \sigma \eqno(22)$$
where $\alpha$ is some positive dimensionless constant. The two
conservation equations now can be integrated to yield the
components of the stress tensor in terms of its trace,
$$\eqalign{\langle T_{++}\rangle~~ &=~~ T^f_{++}~ +~ T^q_{++}~~ =~~
T^f_{++}~~ -~~ \int {{d x^-} \over {\sigma}} \partial_+ (\sigma
T^q_{+-})~~ +~~ A(x^+),\cr \langle T_{--}\rangle~~ &=~~ T^f_{--}~
+~ T^q_{--}~~ =~~ T^f_{--}~~ -~~ \int {{d x^+} \over {\sigma}}
\partial_- (\sigma T^q_{+-} )~~ +~~ B(x^-),\cr}\eqno(23)$$
where $A(x^+)$ and $B(x^-)$ are boundary condition dependent
functions of $x^+$ and $x^-$ respectively. Consider the case of
positive cosmological constant and the observer on the right. A
consistent solution should admit no incoming radiation on ${\cal
I}_R^-$ other than  any matter fields that might be present and
vanish in the absence of the soliton, that is, in the linear
dilaton vacuum.  The stress tensor satisfying these conditions is
$$\eqalign{\langle T_{++}\rangle~~ &=~~ T^f_{++}~ +~ T^q_{++}~~ =~~
T^f_{++}~~ -~~ \alpha \left( {{\partial^2_+ \sigma} \over \sigma}~
-~ {1 \over 2} \left[{{\partial_+ \sigma} \over \sigma}\right]^2
\right)~ -~  {\alpha \over {2 x^{+'2}}}\cr \langle T_{--}\rangle~~
&=~~ T^f_{--}~ +~ T^q_{--}~~ =~~ T^f_{--}~~ -~~ \alpha \left(
{{\partial^2_- \sigma} \over \sigma}~ -~ {1 \over 2}
\left[{{\partial_- \sigma} \over \sigma}\right]^2 \right)~ -~
{\alpha \over {2 x^{-2}}}\cr \langle T_{+-}\rangle ~~ &=~~ T^f_{+-
}~~ +~~ \alpha \partial_+ \partial_- \ln \sigma\cr} \eqno(24)$$
where $x^{+'} = x^+ + 2 \gamma_-/\lambda^2$. It is most convenient
to analyze the above expressions in the coordinate system in which
the metric is manifestly asymptotically flat. Define, therefore the
coordinates $\sigma^\pm = t \pm x$ by
$$\eqalign{x^+~~ =~~ {1 \over \lambda} e^{\lambda \sigma^+}~ -~ {{2
\gamma_-} \over {\lambda^2}}\cr x^-~~ =~~ -~ {1 \over \lambda} e^{-
\lambda \sigma^-}~ -~ {{2 \gamma_+} \over {\lambda^2}}\cr}
\eqno(25)$$
Thus $\sigma^+ \rightarrow \infty$ corresponds to the lightlike
surface $x^+ = \infty$ and $\sigma^- \rightarrow \infty$ to the
lightlike surface $x^- = - \infty$, while $\sigma^- \rightarrow
\infty$ corresponds to the lightlike surface $x^- = - 2 \gamma_- /
\lambda^2$. Transforming the expressions in (23) to the new system,
one finds that
$$\langle T^\sigma_{++}\rangle ~ \rightarrow~ 0,~~~~~~~~ \langle
T^\sigma_{+-}\rangle ~ \rightarrow~ 0 \eqno(26)$$
and
$$\langle T^\sigma_{--}\rangle~~ \rightarrow~~ {{\alpha \lambda^2}
\over 2} \left [ 1~ -~ {1 \over {\left( 1 + {{2 \gamma_+} \over
\lambda} e^{\lambda \sigma^-} \right )^2 }}\right]. \eqno(27)$$
$\langle T^\sigma_{--}\rangle$ is the outgoing flux at ${\cal
I}_R^+$. It grows smoothly from zero at $x^- = - \infty$ to a
maximum of $\alpha \lambda^2 /2$ at $x^- = 2 \gamma_+ / \lambda^2$
on ${\cal I}_R^+$. As expressed in (26) and (27) the quantum stress
tensor represents the Hawking flux from the singularity. It depends
on the soliton mass parameter but not on the mass of the
singularity itself. This would seem to be a general feature of the
Hawking evaporation in this model, having been shown to be true for
the radiation from a black hole formed by an incoming shock wave in
the CGHS model. The integrated flux along ${\cal I}_R^+$ is the
total energy lost by the incoming soliton.  As the flux rapidly
approaches its maximum value of $\alpha \lambda^2 / 2$, the
integrated flux shows an infinite loss of energy if the integral is
performed up to the future horizon at $x^- = 2\gamma_+ /
\lambda^2$. However, as CGHS pointed out, this is a consequence of
having neglected the back reaction and must not be taken seriously.
Instead, one can try to estimate the retarded time, $x^-_\tau$ at
which the integrated Hawking radiation is equal to the mass of the
singularity, $M = 4 \lambda (\Delta_0 - 2 \mu^2/ \lambda^2 + \ln
2)$. One finds
$$\int_{-\infty}^{\sigma^-_\tau} d\sigma^- \langle T^\sigma_{--}
\rangle~~ =~~ {{\alpha\lambda} \over 2} \left[ 1~ -~ {1 \over
{\left( 1 + {{2 \gamma_+} \over \lambda} e^{\lambda \sigma^-_\tau}
\right )}}~ +~ \ln \left(1 + {{2 \gamma_+} \over \lambda}
e^{\lambda \sigma^-_\tau} \right)\right]~~ =~~ M \eqno(28)$$
For a small mass singularity, the retarded time is given by
$$x^-_\tau~~ =~~ -~ {{2 \gamma_+} \over {\lambda^2}}\left( 1~ +~
{{\alpha \lambda} \over M}\right) \eqno(29)$$
which, when na\"\i vely traced backwards, corresponds to the point
$$(x^+_\tau, x^-_\tau)~~ =~~ \left( {{\Delta_0} \over {\gamma_+}}~
+~ {{2 \gamma_-} \over {\lambda^2}} \left[ 1~ +~ {{\alpha \lambda}
\over M} \right],- {{2 \gamma_+} \over {\lambda^2}} \left[ 1~ +~
{{\alpha \lambda} \over M}\right] \right) \eqno(30)$$
on the soliton trajectory. If $x^+_\tau < 0$, the soliton energy
has evaporated earlier than the appearance of its center in the
spacetime. An observer on ${\cal I}_R^+$ sees a white hole which
rapidly radiates away all its energy.  This of course is true only
if the mass of the singularity is small. On the other hand, if
$x^+_\tau$ is greater than zero, the soliton does enter the
spacetime evaporating eventually by $x^-_\tau$ in (29). How
reliable is the estimate of its lifetime above? Assuming that the
soliton center does enter the spacetime before evaporating
completely, the dilaton coupling constant at the turn around point,
$$e^\phi~~ =~~ {1 \over {\sqrt{2 \left(1~ +~ {{\alpha \lambda}
\over M}\right) \left[ {M \over {4\lambda}}~ -~ \ln 2~ -~ {{2\mu^2
\alpha} \over{\lambda M}} \right]}}} \eqno(31)$$
is large for a small mass singularity and signals the breakdown of
the one loop approximation.

On the other hand, if $M$ is large,
$$\int_{-\infty}^{\sigma^-_\tau} d\sigma^- \langle T^\sigma_{--}
\rangle~~ \sim~~ {{\alpha \lambda} \over 2}~ \left[ \ln
{{2\gamma_+} \over \lambda}~ +~ \lambda \sigma^-_\tau \right]~~ =~~
M\eqno(32)$$
or
$$x^-_\tau~~ =~~ -~ {{2\gamma_+} \over {\lambda^2}} \left[1~ +~
e^{-2M/\alpha \lambda} \right] \eqno(33)$$
which, when traced back corresponds to the point
$$(x^+_\tau, x^-_\tau)~~ =~~ \left({{\Delta_0} \over {\gamma_+}}~
+~ {{2 \gamma_-} \over {\lambda^2}} \left[ 1~ +~ e^{-
2M/\alpha\lambda} \right], -~ {{2 \gamma_+} \over {\lambda^2}}
\left[ 1~ +~ e^{-2M/\alpha\lambda} \right]\right)\eqno(34)$$
on the soliton trajectory. The dilaton coupling at this point has
the value
$$e^{\phi}~~ =~~ {1 \over {\sqrt{2 \left( 1~ +~ e^{-2M/\alpha
\lambda} \right) \left( {M \over {4 \lambda}}~ -~ \ln 2~ -~
{{2\mu^2} \over {\lambda}} e^{-2M/\alpha \lambda} \right)}}}
\eqno(35)$$
and is small in the limit of large $M$. The soliton evaporates
completely by the time the observer has reached the event horizon
and the black hole never forms. Moreover this is the limit in which
the one loop approximation is a satisfactory indication of what may
actually be happening.  Similar conclusions can be drawn in the
CGHS (shock wave) model.  In the low mass limit that the dilaton
coupling is large at the turn around point, being proportional only
to $\sqrt{1/\alpha}$ and signalling a breakdown in the one loop
approximation. However, in the large mass limit the dilaton
coupling behaves as the inverse square root of the mass. Thus,
there seems to be no essentially new feature in the evaporation of
the soliton.

Next consider an observer in the left quadrant. As we have
mentioned, because of our choice of $\Delta _0 > 0$ the soliton
center never enters this region and this observer lives in a
universe inhabited only by the tail of the soliton energy and the
singularity described earlier. The appropriate boundary conditions
on the Hawking stress tensor are (a) its vanishing in the absence
of the soliton and (b) no flux across ${\cal I}_L^-$.  It follows
that the quantum contribution to the stress tensor is given by
$$\eqalign{T^q_{++}~~ &=~~ -~~ \alpha \left( {{\partial^2_+ \sigma}
\over \sigma}~ -~ {1 \over 2} \left[{{\partial_+ \sigma} \over
\sigma}\right]^2 \right)~ -~  {\alpha \over {2 x^{+2}}}\cr T^q_{--
}~~ &=~~ -~ \alpha \left( {{\partial^2_- \sigma} \over \sigma}~ -~
{1 \over 2} \left[{{\partial_- \sigma} \over \sigma}\right]^2
\right)~ -~  {\alpha \over {2 x^{-'2}}}\cr T^q_{+-}~~ &=~~ \alpha
\partial_+ \partial_- \ln \sigma \cr} \eqno(36)$$
where $x^{-'} = x^-~ -~ 2 \gamma_+ /\lambda^2$. Going to the system
$\sigma^\pm = t \pm x$ given by
$$\eqalign{x^+~~ &=~~ -~ {1 \over \lambda} e^{- \lambda \sigma^+}~
+~ {{2 \gamma_-} \over {\lambda^2}}\cr x^-~~ &=~~ {1 \over \lambda}
e^{\lambda \sigma^-}~ +~ {{2 \gamma_+} \over {\lambda^2}}\cr}
\eqno(37)$$
in which the metric is manifestly flat at null infinity, ($\sigma^+
\rightarrow \infty$ corresponds to the lightlike line $x^+ = 2
\gamma_-/\lambda^2$ and $\sigma^- \rightarrow -  \infty$ to the
lightlike line $x^+ = 2 \gamma_+ / \lambda^2$ while $\sigma^+
\rightarrow - \infty$ and $\sigma^- \rightarrow \infty$ correspond
to the respective lightlike infinities) one finds
on ${\cal I}_L^+$
$$\langle T^\sigma_{--}\rangle ~ \rightarrow~ 0,~~~~~~~~ \langle
T^\sigma_{+-}\rangle ~ \rightarrow~ 0 \eqno(38)$$
and
$$\langle T^\sigma_{++}\rangle~~ =~~ {{\alpha \lambda^2} \over 2}
\left [ 1~ -~ {1 \over {\left( 1 - {{2 \gamma_-} \over \lambda}
e^{- \lambda \sigma^+} \right )^2 }}\right] \eqno(39)$$
In this part of the world, the radiation is at its maximum value
(of $\alpha \lambda^2 /2$) at early advanced times and decreases to
zero in the far future of ${\cal I}_L^+$. The integrated flux over
any interval is infinite, that is the singularity ``explodes'',
giving up all its energy in a burst and wiping itself out.

We turn now to solitons in 2d gravity with a negative cosmological
constant. We consider the observer in the top quadrangle. In the
absence of ${\cal I}^-$ the only reasonable boundary condition one
may impose upon the stress tensor is that it vanish in the absence
of the soliton, a limit we have defined earlier. The quantum
corrections to the stress tensor then take the form
$$\eqalign{T^q_{++}~~ &=~~ -~~ \alpha \left( {{\partial^2_+ \sigma}
\over \sigma}~ -~ {1 \over 2} \left[{{\partial_+ \sigma} \over
\sigma}\right]^2 \right)~ -~  {\alpha \over {2 x^{+2}}}\cr T^q_{--
}~~ &=~~ -~ \alpha \left( {{\partial^2_- \sigma} \over \sigma}~ -~
{1 \over 2} \left[{{\partial_- \sigma} \over \sigma}\right]^2
\right)~ -~  {\alpha \over {2 x^{-2}}}\cr T^q_{+-}~~ &=~~ \alpha
\partial_+ \partial_- \ln \sigma \cr} \eqno(39)$$
Again, to analyze the tensors it is convenient to go to a system in
which the metric is asymptotically flat: define the system
$\sigma^\pm = t \pm x$ by
$$\eqalign{x^+~~ &=~~ {1 \over \lambda} e^{\lambda \sigma^+}~ -~
{{2 \gamma_-} \over {\lambda^2}}\cr x^-~~ &=~~ {1 \over \lambda}
e^{\lambda \sigma^-}~ +~ {{2 \gamma_+} \over {\lambda^2}}\cr}
\eqno(40)$$
Thus, $\sigma^- \rightarrow - \infty$ corresponds to the lightlike
line $x^- = 2 \gamma_+/\lambda^2$ and $\sigma^+ \rightarrow -
\infty$ to the lightlike line $x^+ = -2 \gamma_- / \lambda^2$ while
$\sigma^\pm \rightarrow \infty$ correspond to the respective
lightlike infinities.

The fluxes across both ${\cal I}_L^+$ and ${\cal I}_R^+$ are now
non-vanishing, each approaching a maximum of $\alpha \lambda^2 /2$
at early times and decreasing steadily in the far future, as $i^0$
is approached. Thus, on ${\cal I}_R^+$, for instance, one finds
$$\langle T^\sigma_{++}\rangle ~ \rightarrow~ 0,~~~~~~~~ \langle
T^\sigma_{+-}\rangle ~ \rightarrow~ 0 \eqno(41)$$
and
$$\langle T^\sigma_{--}\rangle~~ =~~ {{\alpha \lambda^2} \over 2}
\left [ 1~ -~ {1 \over {\left( 1 + {{2 \gamma_+} \over \lambda}
e^{- \lambda \sigma^-} \right )^2 }}\right], \eqno(42)$$
and on ${\cal I}_L^+$
$$\langle T^\sigma_{--}\rangle ~ \rightarrow~ 0,~~~~~~~~ \langle
T^\sigma_{+-}\rangle ~ \rightarrow~ 0 \eqno(43)$$
and
$$\langle T^\sigma_{++}\rangle~~ =~~ {{\alpha \lambda^2} \over 2}
\left [ 1~ -~ {1 \over {\left( 1 - {{2 \gamma_-} \over \lambda}
e^{- \lambda \sigma^-} \right )^2 }}\right]. \eqno(44)$$
The tensors are again independent of the mass of the singularities
but depend on the soliton mass parameter. The integrated flux over
any interval is again infinite because the flux itself approaches
a steady state at early times. Of course this is a consequence of
having neglected the back reaction of the radiation on the
spacetime geometry.

This picture is also similar to that developed in the shock wave
model.${}^{10}$ Even if quantum gravity does permit the formation
of naked singularities, they will evaporate catastrophically
(``explode'') due to the Hawking radiation. To justify this
statement, one must check the validity of the one loop picture by
considering the strength of the dilaton coupling constant at the
point on the soliton center at which the singularities are expected
to detonate. On ${\cal I}_R^+$ this is at the retarded time $x^- =
2\gamma_+/ \lambda^2$ which, when traced back corresponds to the
point $(x_\tau^+,x_\tau^-) = (1/\gamma_+ ( \Delta_0 + 2 \mu^2/
\lambda), 2\gamma_+/\lambda^2)$ on the soliton center and gives for
the dilaton coupling
$$e^\phi~~ =~~ {1 \over {\sqrt{2 \left( {{2\mu^2} \over
{\lambda^2}}~ +~ \Delta_0 \right)}}}. \eqno(45)$$
This is indeed small when the Bondi mass, $M_R$ is large, i.e., the
soliton mass parameter is large. On the other hand, on ${\cal
I}_L^+$ this is at the advanced time $x^+ = - 2\gamma_-/\lambda^2$
which, when traced back, corresponds to the point
$(x_\tau^+,x_\tau^-) = (- 2\gamma_-/\lambda^2, 1/\gamma_- (\Delta_0
- 2 \mu^2/ \lambda))$ on the soliton center and gives for the
dilaton coupling
$$e^\phi~~ =~~ {1 \over {\sqrt{2 \left( {{2\mu^2} \over
{\lambda^2}}~ -~ \Delta_0 \right)}}} \eqno(46)$$
which is again small when the Bondi mass, $M_L$, is large (or the
soliton mass parameter is large).

We shall not discuss here the remaining case, the lower quadrant in
1+1 dilaton gravity with a negative cosmological constant. In this
case an observer is obstructed from reaching future null infinity
by the naked singularity. Unlike the case of the upper quadrant
where the naked singularity forms simultaneously with the emergence
of the soliton center, in the lower quadrant the soliton emerges in
the past. This case may therefore be thought of as a dynamical
model for the ``formation'' of a naked singularity. Because of the
inaccessibility of null infinity to an observer the analysis of the
Hawking radiation requires somewhat different considerations than
the cases we have already discussed. The analysis of this very
interesting case will be reported elsewhere.

In this article we have examined the singularities induced by an
incoming soliton in two dimensional dilaton gravity both with a
positive and negative cosmological constant. The singularities are
neither purely spacelike nor purely timelike but a combination of
the two joined smoothly by lightlike singularities along the
soliton center. The classical stress energy tensor of the Sine-
Gordon field is exponentially vanishing at infinity and the Hawking
radiation is dominant there. We have therefore examined the Hawking
evaporation of these singularities. To do so it was necessary to
impose reasonable boundary conditions on the Hawking tensor.
Arguing that the most sensible conditions are (a) the vanishing of
the tensor in the absence of the soliton energy-momentum and (b)
the absence of incoming radiation on ${\cal I}^-$, we showed that
the essential character of the radiation does not differ
significantly from that produced by an incoming shock wave (the
CGHS model) in the two cases.
\vskip 0.5in

\noindent{\bf Acknowledgements:}

\noindent This work was supported in part by  NATO under contract
number CRG 920096. C. V. acknowledges the partial support of the
{\it Junta Nacional de Investiga\c{c}\~ao Cient\'ifica e
Tecnol\'ogica}, JNICT, Portugal under contract number
CERN/S/FAE/128/94 and L.W. acknowledges the partial support of the
U. S. Department of Energy under contract number DOE-FG02-
84ER40153.
\vfill\eject

\noindent{\bf References:}

{\item{1.}}S. W. Hawking , Comm. Math. Phys. {\bf 43} (1975) 199.

{\item{2.}}J. B. Hartle and S. W. Hawking, Phys. Rev. {\bf D13}
(1976) 2188; B.S. DeWitt, Phys. Rep. 19 (1975) 295; W. Israel,
Phys. Lett. {\bf A 57} (1976) 107; W. G. Unruh, Phys. Rev. {\bf
D14} (1976) 840; S. M. Christensen and S. A. Fulling, Phys. Rev.
{\bf D15} (1977) 2088.

{\item{3.}}James Bardeen, Phys. Rev. Letts. {\bf 46} (1981) 392; D.
N. Page, Phys. Rev. {\bf D25} (1982) 1499; K. W. Howard and P.
Candelas, Phys. Rev. Lett. {\bf 53} (1984) 403; Cenalo Vaz, Phys.
Rev. {\bf D39} (1988) 1776; F. A. Barrios and Cenalo Vaz, Phys.
Rev. {\bf D40} (1989) 1340.

{\item{4.}}S. W. Hawking, Phys. Rev. {\bf D14} (1976) 2460; D. N.
Page, {\it Black Hole Information}, hep-th/9305040; Phys. Rev. {\bf
D9} (1974) 3292; T. Banks, {\it Lectures on Black Holes and
Information Loss}, hep-th/9412131; L. Thorlacius, {\it Black Hole
Evolution}, hep-th/9411020; S. B. Giddings, {\it Quantum Mechanics
of Black Holes}, hep-th/9412138.

{\item{5.}}E. Witten Phys. Rev. {\bf D44} (1991) 314; G. Mandal, M.
M. Sengupta, and S. R. Wadia, Mod. Phys. Letts. {\bf A6} (1991)
1685.

{\item{6.}}C. Callan, S. B. Giddings, J. Harvey and A. Strominger,
Phys. Rev. {\bf D45} (1992) 1005.

{\item{7.}}S. d' Alwis, Phys. Letts. {\bf B289} (1992) 282; Phys.
Rev. {\bf D46} (1992) 5429; A. Billal and C. Callan, Nucl. Phys.
{\bf B394} (1993) 73; J. G. Russo, L. Susskind and L. Thorlacius,
Phys. Rev. {\bf D46} (1993) 3444; Phys. Rev. {\bf D47} (1993) 533;
T. M. Fiola, J. Preskill, A. Strominger and S. Trivedi, {\it Black
Hole Thermodynamics and Information Loss in Two Dimensions}, hep-
th/9403137.

{\item{8.}}J. Polchinski and A. Strominger, Phys. Rev. {\bf D50}
(1994) 7403; hep-th/9407008; A. Strominger, {\it Unitary Rules for
Black Hole Evaporation}, hep-th/9410187.

{\item{9.}}Andrew Strominger, {\it Les Houches Lectures on Black
Holes}, hep-th/9501071

{\item{10.}}Cenalo Vaz and Louis Witten, Phys. Letts. {\bf B325}
(1994) 27.

{\item{11.}}Hak-Soo Shin and Kwang-Sup Soh, {\it Black Hole
Formation by Sine-Gordon Solitons in Two Dimensional Dilaton
Gravity}, hep-th/9501045

{\item{12.}}E. Verdaguer, Phys. Rep. {\bf 229} (1993) 1.

{\item{13.}}S. Weinberg, {\it Gravitation and Cosmology} (John
Wiley \& Sons, Inc. New York, 1972).

{\item{14.}}H. Bondi, M.G.J. Van de Burg and A.W.K. Metzner, Proc.
Roy. Soc. {\bf A269} (1962) 21.

{\item{15.}}R. M. Wald, Comm. Math. Phys. (1977) {\bf 54} 1; Ann.
Phys. (NY) {\bf 110} (1978) 472; Phys. Rev. {\bf D17} (1978) 1477.
\vfill\eject

\noindent{\bf Figure Captions:}

{\item{}}{\bf Figure I:} The Kruskal diagram for $\Lambda > 0$ and
$\Delta_0 > 0$. Regions I \& III are physical ($\sigma > 0)$. A
black hole a white hole and a timelike singularity are joined
smoothly along the soliton center.

{\item{}}{\bf Figure II:} The Kruskal diagram for $\Lambda > 0$ and
$\Delta_0 = 0$. Regions I \& III are physical ($\sigma > 0$). A
black hole and a white hole intersect at the origin.

{\item{}}{\bf Figure III:} The Kruskal diagram for $\Lambda < 0$
and $\Delta_0 > 0$. Regions II \& IV are physical ($\sigma > 0)$.
Two timelike  singularities are joined smoothly at the soliton
center.

{\item{}}{\bf Figure IV:} The Kruskal diagram for $\Lambda < 0$ and
$\Delta_0 = 0$. Regions II \& IV are physical ($\sigma > 0)$. Two
timelike  singularities intersect at the origin.
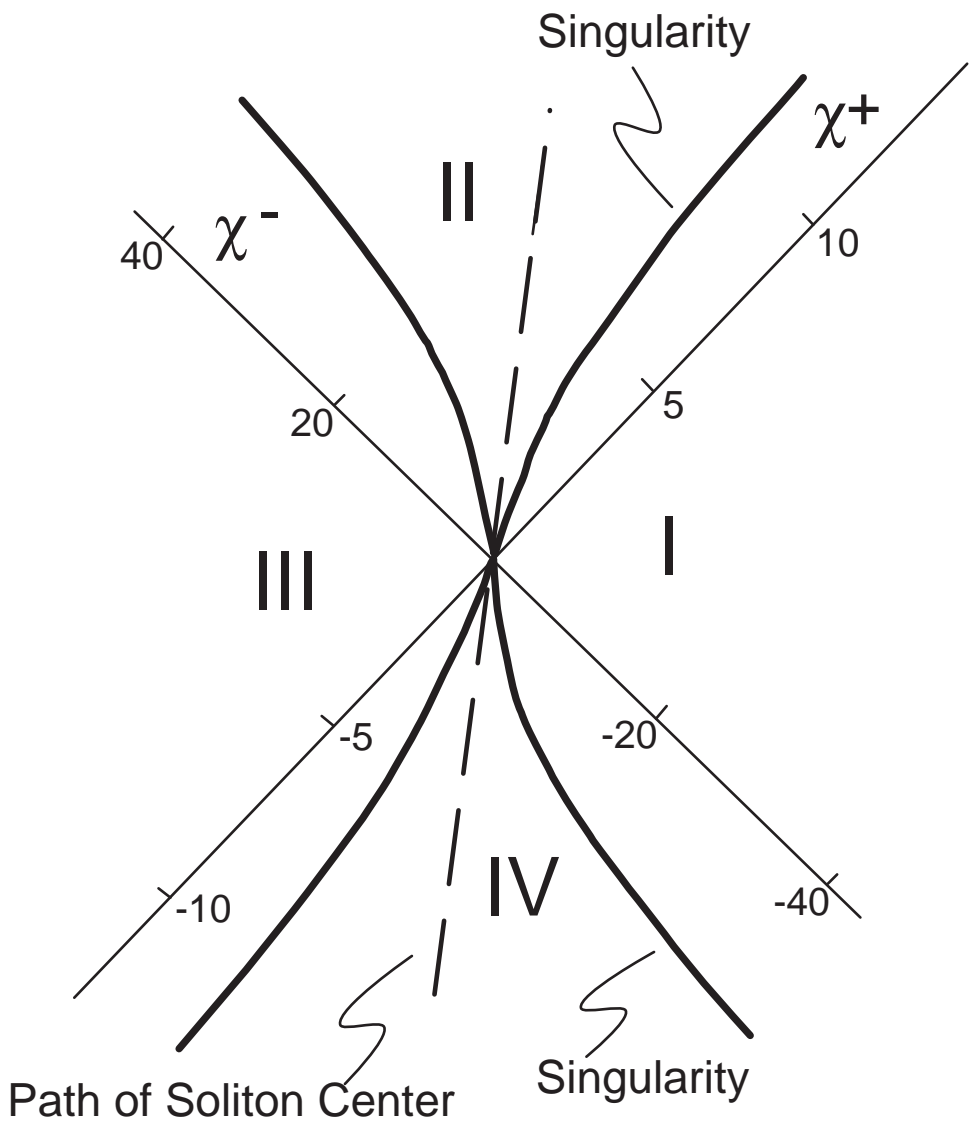
\end